\documentclass[twocolumn,aps,unsortedaddress]{revtex4-2}

\usepackage{amsmath}
\usepackage{mathtools}
\usepackage{bm}
\usepackage{amssymb}
\usepackage{svg}
\usepackage[colorlinks=true,linkcolor=blue,citecolor=blue,urlcolor=blue]{hyperref}
\usepackage[normalem]{ulem}
\usepackage{svg-extract}

\newcommand{\Tr}[1]{\mathrm{Tr}{\left\{#1\right\}}}
\newcommand{\Ropt}[1]{{#1}^{\mathrm{R}}}
\newcommand{\Aopt}[1]{{#1}^{\mathrm{A}}}

\ifx\pdfoutput\undefined
\usepackage[dvipdfmx]{graphicx}
\usepackage[dvipdfmx]{hyperref}
\usepackage[dvipdfmx]{color}
\usepackage[dvipdfmx]{xcolor}
\else
\usepackage{graphicx}
\usepackage{hyperref}
\usepackage{color}
\usepackage{xcolor}
\fi

\usepackage{ulem}
\usepackage{comment}
\usepackage{txfonts}
\usepackage{braket}

\DeclarePairedDelimiter\abs{\lvert}{\rvert}%
\DeclarePairedDelimiter\norm{\lVert}{\rVert}%

\makeatletter
\let\oldabs\abs
\def\abs{\@ifstar{\oldabs}{\oldabs*}}
\let\oldnorm\norm
\def\norm{\@ifstar{\oldnorm}{\oldnorm*}}
\makeatother

\begin{document}

\title{
%Current carrying boundary states inspired by the inverse spin Hall effect in three-dimensional Dirac fermions\\
%Exploring Anomalies and Inverse Spin Hall Effects in Dirac Semimetals with Spatially Varying Chiral Gauge Fields\\
% Inverse Spin Hall Effect in Dirac Semimetals and Quantum Anomalies\\
% Inverse Spin Hall Effect in Dirac Semimetals Induced
% by Chiral Gauge fields \\
%Nonequilibrium Imbalance of Surface and Anomalous flows:
%Invserse Spin Hall effect in Dirac Semimetals
% \\
%Nonequilibrium Imbalance of Surface and Anomalous flows:\\
Inverse Spin Hall Effect in Nonequilibrium Dirac Systems
Induced by Anomalous Flow Imbalance
}

\author{Hung-Hsuan Teh$^1$}
\email{teh@issp.u-tokyo.ac.jp}
\author{Tokiro Numasawa$^1$}
\author{Shun Okumura$^{1,2}$}
\author{Takashi Oka$^1$}
\email{oka@issp.u-tokyo.ac.jp}
\affiliation{
$^1$The Institute of Solid State Physics, The University of Tokyo, Kashiwa, Chiba 277-8581, Japan\\
$^2$Deperment of Applied Physics, The Univerity of Tokyo, Hongo, Tokyo, 113-8656, Japan
}

\date{\today}

\begin{abstract}
% Motivated by a recent experiment on the inverse spin Hall effect in bismuth thin films, we study the surface states of three dimensional Dirac electrons in the presence of a localized chiral gauge field and thermodynamic gradients. 
% performing 
% a tight binding model calculation
% using nonequilibrium green's functions
We study Dirac fermions in the presence of a space-dependent chiral gauge field and thermodynamic gradients, establishing a connection to the inverse spin Hall effect. The chiral gauge field induces a chiral magnetic field, resulting in a surface Fermi arc state and a chiral Landau level state which, although is delocalized in the bulk, we show to be more robust against impurities. By applying chemical potential and temperature gradients, we achieve nonzero charge currents, with each gradient leading to distinct Fermi level dependencies, both of which have been observed in a recent experiment. Unlike the conventional mixed axial-gravitational anomaly, our currents require a noncollinear chiral magnetic field and thermodynamic gradient. We further derive low-energy transport formulas and demonstrate the importance of carefully treating the ultraviolet cutoff for understanding our lattice calculations.
\end{abstract}

\maketitle

% Bulk Anomalous current by chiral Landau level:
% $$J_y=\frac{e^2}{h}B^5_y\mu_{\rm cLL},\;J^e_y=\frac{e^2}{h}B^5_y\frac{1}{12}T^2_{\rm cLL}$$

% Surface Anomalous current by Fermi arc:
% $$J_y=-\frac{e^2}{h}B^5_y\mu_{\rm FA},\;
% \;J^e_y=-\frac{e^2}{h}B^5_y\frac{1}{12}T^2_{\rm FA}$$

% Imbalance of Anomalous currents in nonequilibrium: 
% $$J_y=\frac{e^2}{h}B^5_y(\mu_{\rm cLL}-\mu_{\rm FA}),$$
% $$J^e_y=\frac{e^2}{h}B^5_y\frac{1}{12}(T^2_{\rm cLL}-T^2_{\rm FA})$$

% Relation between "Spin current" and chiral magnetic field $B^5$: 
% $$J_z^5=\alpha  A^5_z $$
% $$\nabla_x S^z=\nabla_x J_z^5=\alpha \nabla_x A^5_z$$
% This relation is only true at linear order. 

% Inverse Spin Hall effect: 
% $$J_y=\frac{e^2}{h}\frac{1}{\alpha}\nabla_x S^z(\mu_{\rm cLL}-\mu_{\rm FA})$$
% $$J_y^e=\frac{e^2}{h}\frac{1}{\alpha}\nabla_x S^z\frac{1}{12}(T^2_{\rm cLL}-T^2_{\rm FA})$$

%skew the balance between Fermi surfaces
\textit{Introduction} --- Quantum materials respond to external forces, exhibiting currents that are governed by symmetry properties and conservation laws. The presence of multiple symmetries can lead to intriguing cross-correlations among these currents. This has profound implications in the realm of spintronics where the interconnection between charge and spin currents has attracted attention, particularly in materials where electron spin orientation exhibits weak conservation. A prime illustration of this is the spin Hall effect, where an applied electric field induces not only a direct charge current but also a transverse spin current, showing a fundamental spin-charge interplay~\cite{DYAKONOV1971459,HirschSHE,SZhangSHE,SinovaISHE,MurakamiScience}.
%Quantum materials' currents, governed by symmetry and conservation laws, show intriguing cross-correlations, impacting spintronics. The spin Hall effect illustrates electric fields inducing charge and transverse spin currents.

Conversely, the {\it inverse} spin Hall effect (ISHE) represents a less explored frontier, initially observed at the interface between a ferromagnetic electrode and platinum~\cite{SaitohISHE}. 
At the interface, a spin current injected into platinum results in a transverse charge current,
underscoring the reciprocal nature of spin-charge conversion. Recently, a helicity-dependent transverse photo-excited current in bismuth based Dirac semimetals was measured~\cite{kawaguchi2020giant}. This current distinguishes itself from conventional photocurrents, which typically flow parallel to the plane of incidence. Although there is no ferromagnetic electrode in the system, its origin is speculated to be the inverse spin Hall effect. In this scenario, circularly polarized light induces spin accumulation at the irradiated surface. This mirrors the magnetic proximity effect originally achieved with the ferromagnet. 
The sign and size of the photo-excited current depend sensitively on the electron density and its microscopic understanding is still an open issue.

\begin{figure}
\centering
\includegraphics[width=0.47\textwidth]{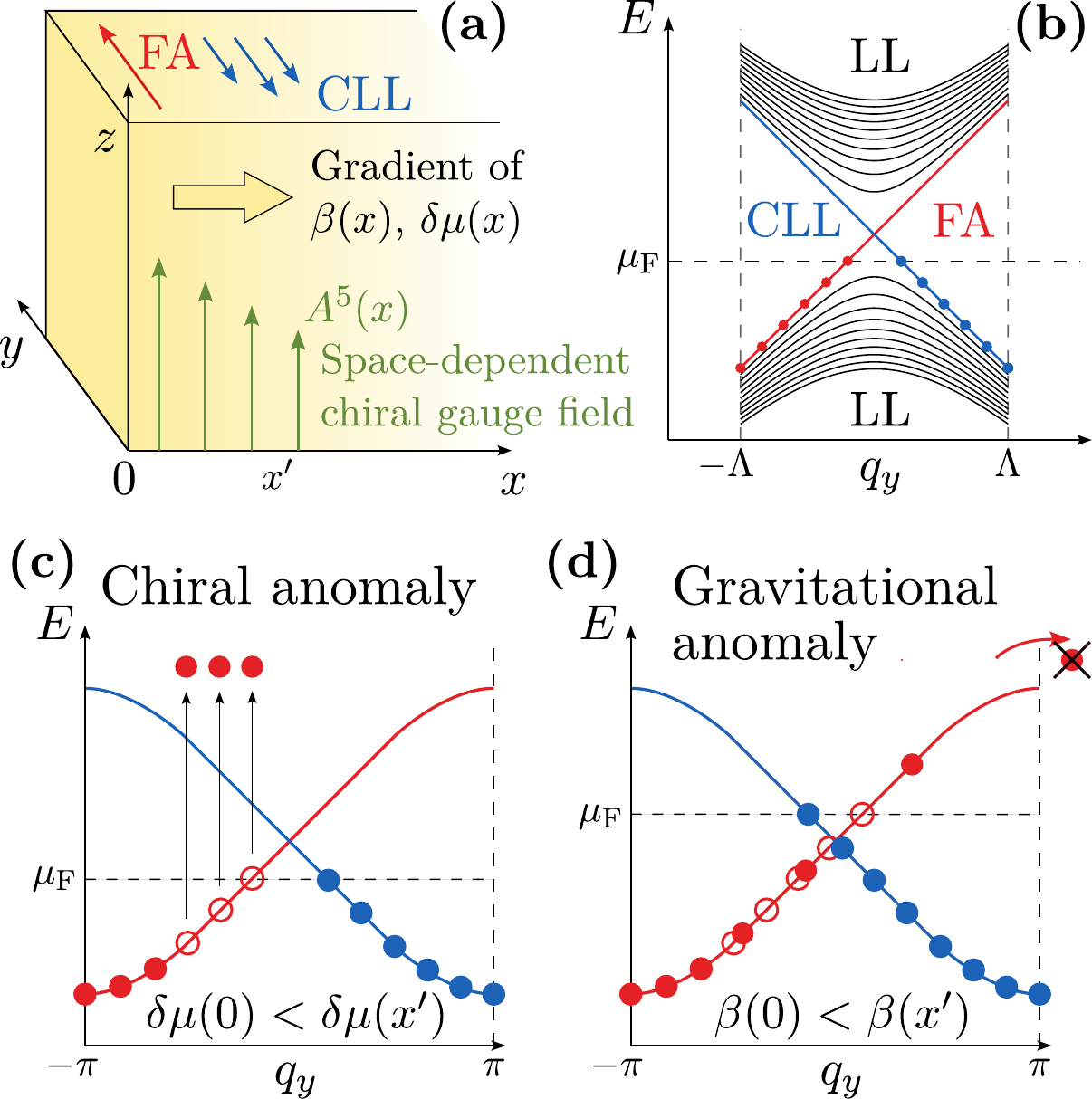}
\caption{(a) Key components for achieving a finite anomalous current along the $y$ direction at one boundary. $x'$ denotes the position where the CLL state is located at.
%The background chiral gauge field $\mathbf{A}^{5}_{z}(x)$ (green arrows) turns the 3D Dirac semimetal (yellow block) into a Weyl semimetal. Two topologically protected chiral states emerge --- one is a conventional surface Fermi arc state (red arrow), and the other is a chiral Landau level state (blue arrows) resulting from the space dependence of the chiral gauge field. A gradient of the inverse temperature $\beta$ or the chemical potential $\delta\mu$ along the $x$ direction (yellow arrow) is used for generating an imbalance between the two chiral states, leading to a net current.
(b) Dispersion of the FA, CLL, and LL states in the presence of chiral magnetic field. $\Lambda$ is a UV cutoff parameter below which the continuum Dirac model in Eq.~\eqref{eq:bloch_H_dirac_A5} becomes a good description. (c) Chiral anomaly-like scenario: The imbalance of the electron population is generated by the chemical potential gradient $\delta\mu(x)$. (d) Gravitational anomaly-like scenario: The temperature gradient $\beta(x)$ leads to different spreads of electrons. The boundary of Brillouin zone gives rise to the distinct current contributions from the two chiral states --- part of the FA electron does not contribute to the current. The corresponding transport formulas for the two scenarios are derived in Eqs.~\eqref{eq:Jy_mbg_general}-\eqref{eq:Jy_beta_grad}.
}\label{fig:fig1}
%\caption{Schematic pictures of two setups where the effects of circularly polarized lights (CPLs), as shown in green spiral curves, are equivalent to chiral gauge fields as shown in green arrows: (a) The CPL propagates in the normal direction to the Dirac semimal surface ($x$-$y$ plane), and penetrates through the semimetal without a decay. The resulting chiral gauge fields, which is constant fields pointing in the $z$ direction, lead to an emergence of Fermi arc states, as shown in red and blue, flowing around surfaces perpendicular to the $x$-$y$ plane. (b) The CPL illuminates on the Dirac semimetal with an incident angle $\phi$. We further consider the skin effect such that the CPL decays as it passes through the semimetal, and therefore the corresponding chiral gauge field also decays. Distinct from (a), one of the Fermi arc states leaks into the bulk, becoming a chiral Landau level state.
\end{figure}

\textit{From ISHE to Dirac Systems ---} In this paper we study a simplified model of current generation inspired by the ISHE scenario described in Ref.~\cite{kawaguchi2020giant}. Concepts in spintronics are translated into the framework of Dirac and Weyl fermions. Note that a 3D Dirac fermion is composed of two Weyl fermions with left and right chiralities.
%with the $U_5(1)$ chiral charge representing the difference between Weyl charges.
The chiral density, i.e. the population difference between the two Weyl fermions, is weakly conserved in bismuth due to a small mass term that mixes the two Weyl fermions. This situation is analogous to spintronics, where spin-orbit coupling causes the total number of spin to be non-conserved.\\
%We identify spin $\mathbf{S}$ to chiral gauge field $\mathbf{A}^{5}$ with some proportional constant. This identification is consistent with the fact that magnetization induces a chiral gauge field $\mathbf{A}^{5}$~\cite{PhysRevB.94.115312}. 
The magnetic proximity effect in spintronics is then translated into a localized background chiral gauge field (CGF) $\mathbf{A}^{5}(\mathbf{x})$ spatially varying. The system under consideration is illustrated in Fig.~\ref{fig:fig1}(a), extending to $x>0$ with the CGF localized near the interface at $x=0$ and pointing in the $z$ direction. We interpret the change in the background CGF as the ``spin current'' running in the $x$ direction. Notice that this interpretation is valid even for spin nonconserving processes.
% The role of spin current is played by the chiral current, 
% and particular focus is given to the anomaly-like transport~\cite{CHERNODUB20221,Arakiconversion}. To this end, it is important to note that a 3D Dirac fermion is composed of two Weyl fermions with left and right chiralities. 
%, with the $U_5(1)$ chiral charge representing the sum of charges from both Weyl fermions, but with opposite signs. 

An additional critical component for generating a charge current is a counterpart to the ``dissipative spin current" discussed in Ref.~\cite{SaitohISHE}. We interpret this as a dissipative flow toward the bulk, represented by a gradient in certain thermodynamic quantities. Specifically, we consider two candidates: (i) chemical potential $\mu(x)$, and (ii) inverse temperature $\beta(x)$ that alter in the $x$-direction.
Below, we demonstrate that the gradients of these two thermodynamic quantities lead to charge currents in the $y$ direction, each with a distinct dependency on fermionic density. One reason for selecting these two quantities is that they give rise to transport phenomena in Weyl semimetals induced by (i) the chiral anomaly and (ii) the gravitational anomaly, respectively~\cite{landsteiner2016notes,PhysRevB.86.115133,burkov2015chiral,PhysRevB.99.140201,CHERNODUB20221}.

It is natural to conceive that a chiral magnetic field $\mathbf{B}^{5}$ induced by the CGF, $\mathbf{B}^{5}=\nabla\times\mathbf{A}^{5}$, generates a charge current in the bulk according to the chiral magnetic effect (CME)~\cite{PhysRevD.78.074033,KHARZEEV2014133}. In our setup with $\mathbf{B}^5=(0,B^5_y,0)$, this charge current runs in the $y$ direction and CME gives $J_{y}=B^{5}_{y}\mu_\mathrm{CLL}/2\pi^{2}$, where $\mu_\mathrm{CLL}$ is the chemical potential of the bulk state. However, this is only half of the story. There exists another current contribution from the boundary state, and the thermodynamic gradients lead to an imbalance between the two contributions. In addition, the CME expression does not account for temperature dependence. As we will show, the total charge current has the following general form,
\begin{align*}
J_{y}=\frac{1}{2\pi^{2}}\alpha(\mu,\beta,\Lambda)B^{5}_{y},
\end{align*}
where the coefficient $\alpha$ involves the chemical potential $\mu$ and the inverse temperature $\beta$ around the bulk and boundary states, and an ultraviolet (UV) cutoff $\Lambda$ as well (see Eq.~\eqref{eq:Jy_mbg_general} for the result). A similar expression is derived for the total energy current (see \cite{SM} for details).

% In this paper, we show that a charge current is generated as 
% $$J_y=\alpha(\mu,T,\Lambda)\nabla_x A^5_z$$
% $\nabla_xA^5_z$ is related to the spin current $\nabla_x J^5_z$

\textit{Dirac Hamiltonian and Chiral Gauge Field} --- We consider the general Dirac Hamiltonian with a background CGF,
\begin{align}\label{eq:bloch_H_dirac_A5}
H=v\sum_{j}\gamma^{0}\gamma^{j}q_{j}+m\gamma^{0}
+\sum_{j}A^{5}_{j}\gamma^{0}\gamma^{j}\gamma^{5},
\end{align}
where the Latin index $j$ runs over the three space coordinates and $q_{j}$ is the corresponding momentum. $v$ denotes the Fermi velocity, $m$ labels the Dirac mass, and $A^{5}_{j}$ is the strength of the background CGF in the $j$-th direction. $\{\gamma^{\mu}\}$ represent the standard gamma matrices satisfying Clifford algebra, with the Greek index $\mu$ labeling the spacetime coordinates. In this paper, we apply the chiral representation for the gamma matrices, namely $\gamma^{0}=[0\,I;\,I\,0]$ and $\gamma^{j}=[0\,\sigma^{j};\,-\sigma^{j}\,0]$ where $I$ and $\sigma^{j}$ denote the identity and the $j$-th component of Pauli matrices.

Microscopically, there are several origins for the CGF: (i) Spin polarization induced by magnetic electrodes or surface magnetization through irradiating a circularly polarized laser~\cite{PhysRevB.94.115312}, (ii) Floquet-induced CGF~\cite{PhysRevB.93.155107,yoshikawa2022lightinducedchiralgaugefield}, and (iii) lattice distortion~\cite{PhysRevLett.115.177202,PhysRevX.6.041021}. When the system is homogeneous and the CGF is large enough, i.e. $\abs{\mathbf{A}^{5}}>m$, the Dirac cone splits into a pair of Weyl cones due to the breakdown of time-reversal symmetry, and Fermi arc (FA) states appear on the surfaces. The Weyl nodes are topologically protected, and various microscopic models for Weyl semimetals have been proposed in prior research~\cite{PhysRevB.83.205101,PhysRevLett.107.127205,RevModPhys.90.015001}, with their transports also being intensively studied~\cite{PhysRevB.86.115133,KHARZEEV2014133,PhysRevB.87.235306,PhysRevB.88.245107,PhysRevB.88.104412,PhysRevLett.111.027201,PhysRevX.4.031035,PhysRevB.94.245121,PhysRevB.87.161107,PhysRevLett.108.046602,PhysRevB.90.165115,lucas2016hydrodynamic}.
%In the presence of the chiral gauge field, when $m<\abs{\mathbf{A}^{5}}$, the Dirac cone splits into a pair of Weyl cones due to the break of time-reversal symmetry, and the Fermi arc states appear on the surface. The Weyl nodes are topologically protected and various microscopic models for Weyl semimetal have been proposed in prior research~\cite{PhysRevB.83.205101,PhysRevLett.107.127205,RevModPhys.90.015001}. The background chiral gauge field can be generated by lattice distortion~\cite{PhysRevLett.115.177202,PhysRevX.6.041021}, magnetization~\cite{PhysRevB.94.115312} and/or circularly polarized light (CPL)~\cite{PhysRevB.93.155107} in materials.
%Take the CPL case as an example: the light beam propagates in the $z$ direction with a frequency $\Omega$ and shines on the Dirac semimetal, which is described by the first two terms of Eq. (\ref{eq:bloch_H_dirac_A5}). The effect of the light field can be incorporated by employing the minimal substitution $q_{j}\rightarrow q_{j}+A_{j}$, where $\mathbf{A}(t)=A[\sin{\Omega t},\cos{\Omega t},0]$ represents the vector gauge field of the CPL with a field strength $A$. The time-dependent Hamiltonian then can be further simplified by applying the high-frequency expansion, which is commonly used in Floquet physics~\cite{oka2019floquet}, and the leading order terms of effective Hamiltonian become Eq. (\ref{eq:bloch_H_dirac_A5}) with $A^{5}_{z}=(vA)^{2}/\Omega$.

Inspired by experiments, we also allow in the CGF for a space dependency. %which reflects for instance the inherent skin effect of the circularly polarized laser. 
This space dependency leads to an effective chiral magnetic field, $\mathbf{B}^{5}=(0,B^{5}_{y},0)$ in our setup, which further induces Landau levels (LLs) and a chiral Landau level (CLL) as shown in Fig.~\ref{fig:fig1}(b)~\cite{PhysRevX.6.041046}. As we will show, an additional thermodynamic gradient creates an imbalance between the two chiral states as schematically summarized in Figs.~\ref{fig:fig1}(c) and \ref{fig:fig1}(d). In contrast to the mixed axial–gravitational anomaly, which requires collinear (chiral) magnetic field and thermal gradient for obtaining a nonzero charge current~\cite{PhysRevLett.107.021601,gooth2017experimental,PhysRevResearch.2.013088}, our thermal gradient needs to point away from the chiral magnetic field.
%, which can reflect the inherent skin effect of the circularly polarized laser --- decaying as it penetrates materials. This space dependency can alternatively be realized when a temperature gradient is present in ferromagnetic materials, causing magnetization to disappear in the region of elevated temperature.

\textit{Lattice Model} --- To study edge properties of Eq.~\eqref{eq:bloch_H_dirac_A5} and calculate current, we construct the following lattice model~\cite{doi:10.1126/science.1133734},
\begin{align}\label{eq:lattice_H_dirac_A5}
H=\sum_{l=1}^{L}\Bigg\{
&\left[
v\sum_{j=y,z}\gamma^{0}\gamma^{j}\sin{(q_{j}+A_{j})}+M(\mathbf{q})\gamma^{0}
\right]c_{l}^{\dagger}c_{l}\notag\\
&+v\gamma^{0}\gamma^{x}\frac{1}{2i}\left(
e^{iA_{x}}c_{l}^{\dagger}c_{l+1}-e^{-iA_{x}}c_{l}^{\dagger}c_{l-1}\right)\notag\\
&-\lambda\gamma^{0}\left(e^{iA_{x}}c_{l}^{\dagger}c_{l+1}+e^{-iA_{x}}c_{l}^{\dagger}c_{l-1}\right)\notag\\
&+A^{5}_{z}(x)\gamma^{0}\gamma^{z}\gamma^{5}c_{l}^{\dagger}c_{l}
\Bigg\},
\end{align}
where $M(\mathbf{q})=m+6\lambda-2\lambda[\cos{(q_{y}+A_{y})}+\cos{(q_{z}+A_{z})}]$, and $v$ represents the Fermi velocity. We implement open boundaries in the $x$ direction where the lattice site is labeled by $l$, and there are $L$ sites in total. %Here, we set the lattice constant $a=1$; hence, the coordinate $x$ corresponds to discretized $l$ in this lattice model.
We introduce the electron hopping integral $\lambda$ to gap out all the degeneracies at the boundary of Brillouin zone. We intentionally include the vector gauge field $A_{j}$ so that we can calculate the current operator later. Without loss of generality, we restrict the CGF to pointing in the $z$ direction. The Dirac Hamiltonian in Eq.~\eqref{eq:bloch_H_dirac_A5} is recovered in the small $q_{j}$ limit.
%under a periodic boundary condition in the $x$ direction.
Hereafter our primary focus is on the massless situation since the results change only quantitatively when $m$ is small (see Fig.~\ref{fig:fig3}(d) for the mass dependency).

\begin{figure}
\centering
\includegraphics[width=0.47\textwidth]{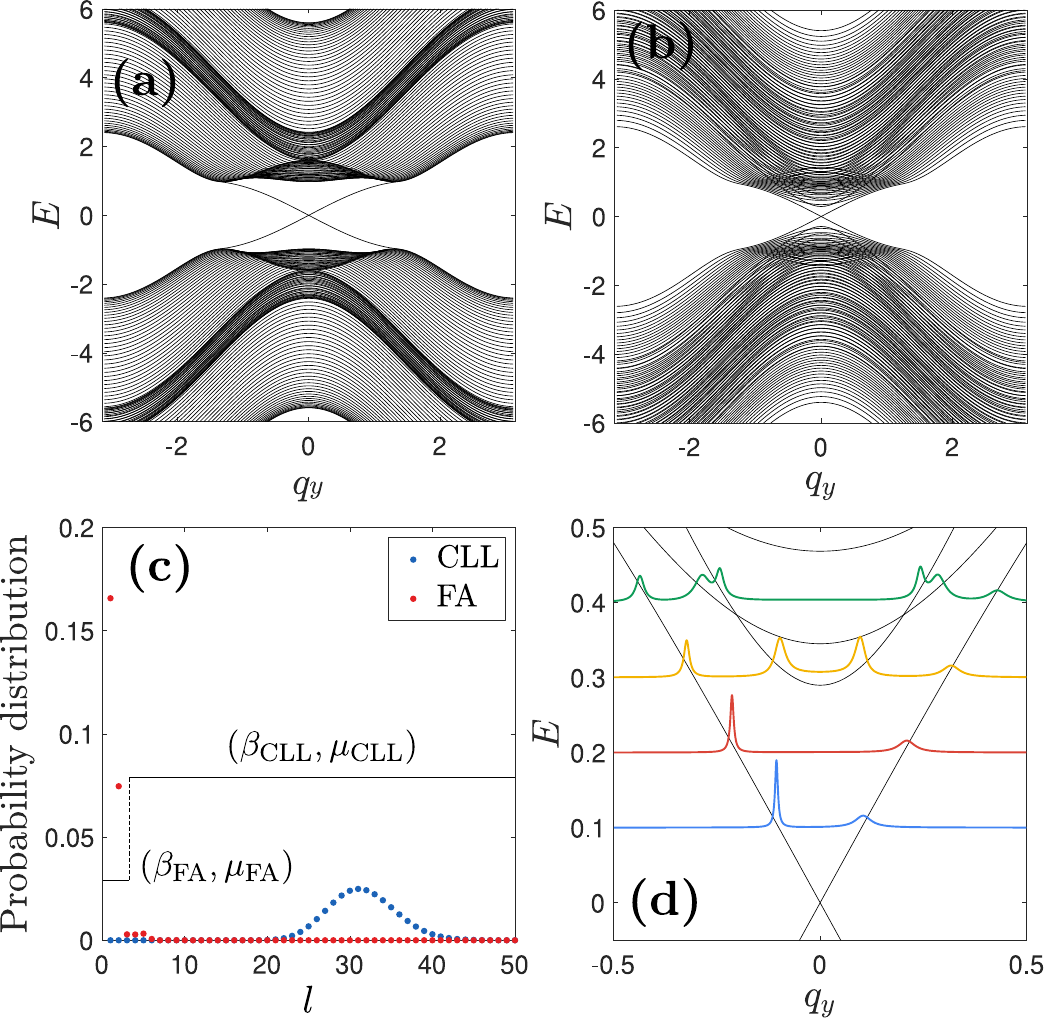}
\caption{(a) Band structures of the lattice Hamiltonian in Eq.~\eqref{eq:lattice_H_dirac_A5} with a constant CGF $A^{5}$, and (b) with a gradient CGF $A_z^{5}(x)$. Parameters used in the calculations are: $\Omega=0.1$, $v=1$, $m=0$, $\lambda=-1$, $L=\xi=50$, and $A^{5}=1.6$. The band structure in (b) has the same chiral state eigenenergies as in (a), but involves more intricate LL structures. (c) Real space probability densities of the two chiral states obtained in (b) at $q_{y}=0.8$. When the CGF becomes space-dependent, one FA state stays localized at the boundary (red dots), and the other turns into a CLL state, delocalizing in the bulk (blue dots). Black lines are effective inverse temperatures and chemical potentials considered in the derivation of Eq.~\eqref{eq:Jy_mbg_general}. 
(d) Calculations for spectral function (color lines) at different energies of the spectrum in Fig.~\ref{fig:fig1}(b). The CLL state remains more robust against the scattering than the FA state. The impurity strength $\kappa=0.02$.
\label{fig:fig2}}
\end{figure}

\begin{figure}
\centering
\includegraphics[width=0.47\textwidth]{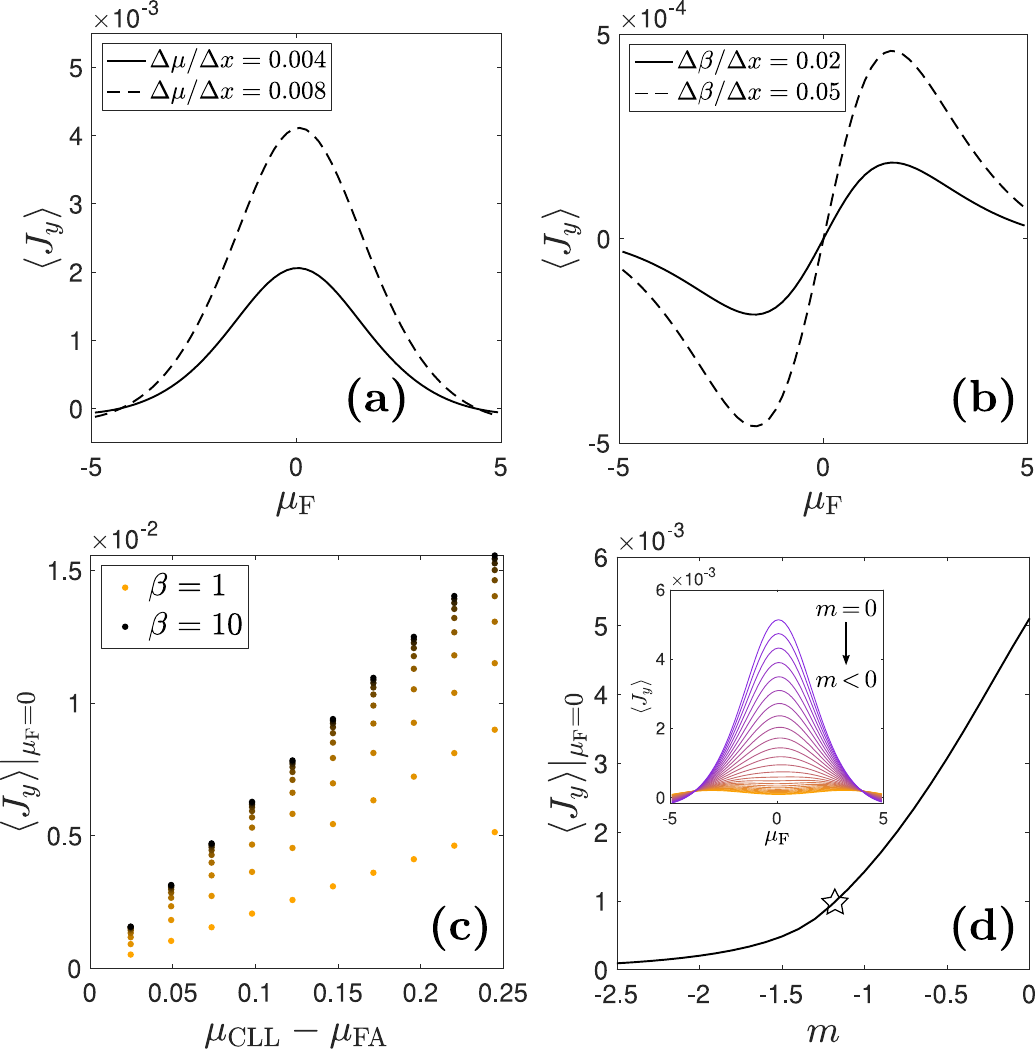}
\caption{(a)(b) Numerical calculations for the anomalous current as a function of Fermi level $\mu_{\text{F}}$ when (a) a chemical potential gradient is applied at a fixed $\beta=1$, and (b) when a temperature gradient is included ($\beta_{0}=1$). The current results show even and odd symmetry under different thermodynamic gradients. Other parameters are the same as those used in Fig.~\ref{fig:fig2}(b). (c) Temperature dependency of $\langle J_{y}\rangle$ as a function of $\mu_{\text{CLL}}-\mu_{\text{FA}}$ in the chiral anomaly-like scenario. Ten values of $\beta=1,2,...,10$ are included with different colors (from light to dark). The linear dependence at each temperature and the asymptotic behavior approaching the low temperature limit are consistent with Eq.~\eqref{eq:Jy_mu_grad}. (d) Mass dependency of $\langle J_{y}\rangle$. Gap opens around $m=-1.2$ labeled by the star. The current becomes smaller when $\lvert m\rvert$ is larger, but remains finite even in the trivial phase
.\label{fig:fig3}}
\end{figure}

\textit{Chiral States} --- In Fig.~\ref{fig:fig2}(a), we show the eigenenergies obtained by exact diagonalization for Eq.~\eqref{eq:lattice_H_dirac_A5} when $A^{5}_{z}(x)=A^5$ is \textit{constant}. As expected, two FA states emerge and localize at $x=0$ and $x=L$. The localized wave function $\psi_{\text{FA}}(x)$ can be obtained by solving Schr\"{o}dinger equation for the continuum model of Eq.~\eqref{eq:bloch_H_dirac_A5} (replacing $q_{x}$ with $-i\partial_{x}$). Near the boundary $x=0$, specifically for $A^{5}>0$, the wave function takes the form~\cite{PhysRevB.88.245107},
\begin{align}
\psi_{\text{FA}}\sim
\begin{bmatrix}
e^{(q_{z}-A^{5}/v)x}c_{\text{L}}\\
e^{(q_{z}-A^{5}/v)x}(-ic_{\text{L}})\\
e^{(-q_{z}-A^{5}/v)x}c_{\text{R}}\\
e^{(-q_{z}-A^{5}/v)x}ic_{\text{R}}
\end{bmatrix},
\end{align}
and the corresponding eigenenergy is $E=vq_{y}$. The normalization coefficients $c_{\text{L},\text{R}}$ satisfy $c_{\text{R}}=\pm ic_{\text{L}}$ due to the particle-hole symmetry of Eq.~\eqref{eq:lattice_H_dirac_A5}, $\gamma^{x}H(q_{y},q_{z})(\gamma^{x})^{\dagger}=-H(-q_{y},-q_{z})$ at $\mathbf{A}=0$, and the sign $\pm$ is determined by the parameters used in the lattice Hamiltonian. A normalizable solution exists only when $-A^{5}<vq_{z}<A^{5}$.
%Note that the delocalization length depends on $A^{5}_{z}$ as well as $q_{z}$. Also 
Note that the Hermiticity condition derived at the boundary is satisfied for this solution~\cite{10.1093/ptep/ptx053}, $\psi_{\text{FA}}^{\dagger}\gamma^{0}\gamma^{1}\psi_{\text{FA}}\vert_{x=0}=0$, indicating that a net current running in the $x$ direction is forbidden. The other FA state can be solved in a similar way and it acquires an eigenenergy $E=-vq_{y}$. The emergence of these FA states localized at the two boundaries requires a homogeneous background CGF throughout the entire sample. 

By contrast, in the following we consider a localized $A^{5}_{z}$ field leading to two topologically protected chiral states as well but only one boundary is needed. %Since the two linear dispersions have opposite slopes, chiral anomalies cannot arise without applying external fields.
We set $A^{5}_{z}(x)=A^{5}(1-x/\xi)\Theta(\xi-x)$ where $\Theta$ is the Heaviside step function and $\xi$ is a cutoff. In Fig.~\ref{fig:fig2}(b), we show the resulting spectrum which shares the same linear dispersion $E=\pm vq_{y}$ near $q_{y}=0$ as in Fig.~\ref{fig:fig2}(a). However, there are two differences: First, the bulk spectrum in Fig.~\ref{fig:fig2}(b) has additional LL structures. Second, one of the chiral states in Fig.~\ref{fig:fig2}(a) is no longer a FA state. As shown in Fig.~\ref{fig:fig2}(c), it leaks into the bulk and becomes a CLL state. %Note that if the chiral gauge field has component and space dependency solely in one direction the chiral magnetic field vanishes and so the two Fermi arc states stay localized on the boundaries, which is in line with the experimental observation.\\
Both differences can be attributed to the effective chiral magnetic field $\mathbf{B}^{5}$. We use the canonical quantization to solve Schr\"{o}dinger equation in the presence of $\mathbf{B}^{5}$, and obtain the CLL wavefunction (see \cite{SM} for details),
\begin{align}
\psi_{\text{CLL}}\sim
\begin{bmatrix}
e^{-A^{5}x^{2}/2v\xi+(-q_{z}+A^{5}/v)x} (-c_{\text{L}})\\
e^{-A^{5}x^{2}/2v\xi+(-q_{z}+A^{5}/v)x} (-ic_{\text{L}})\\
e^{-A^{5}x^{2}/2v\xi+(q_{z}+A^{5}/v)x} (ic_{\text{R}})\\
e^{-A^{5}x^{2}/2v\xi+(q_{z}+A^{5}/v)x} c_{\text{R}}
\end{bmatrix},\label{eq:cll_state}
\end{align}
with an eigenenergy $E_{0}=-vq_{y}$ unchanged from the FA state. The normalization coefficients $c_{\text{L}}$ and $c_{\text{R}}$ satisfy $c_{\text{R}}=\pm ic_{\text{L}}$ due to the particle-hole symmetry. The sign $\pm$ also hinges on the parameters of the Hamiltonian. The delocalized Gaussian characteristic of the CLL state raises concerns that it might be strongly destroyed by impurities.
%Surprisingly, this CLL state is still topologically protected which can be understood by evaluating $\int dx\,\psi_{\text{FA}}^{*}H\psi_{\text{CLL}}=0$ (More precisely, when the edge at $x=L$ is considered, $\int dx\,\psi_{\text{FA}}^{*}\psi_{\text{CLL}}\sim e^{-Lx}$).
%In both Fig. \ref{fig:fig2} (a) and (b) since the FA and CLL states have the linear dispersions with opposite slopes, we do not expect to see chiral anomalies without applying any external fields.

%In order to achieve an anomalous finite current in the $y$ direction, we consider the three scenarios described in Fig. \ref{fig:fig1} (c)-(e). Let us start with the non-Hermitian case. We can utilize the significant difference of real space profiles for the FA and CLL states --- they response to impurities distinctly.

\textit{Robustness of CLL} --- We check the robustness of the CLL state by considering disorders through a self-energy under the first Born approximation~\cite{bruus2004many}, $\Sigma^{\text{FBA}}(\omega)=\int dq_{y}dq_{z}\,1/\left[\omega+i\eta-H(q_{y},q_{z})\right]$ at frequency $\omega$. For simplicity, we require the impurities to be much heavier than the electron so that the electron is scattered locally, and we assume that the scattering is independent on orbitals labeled by $\nu$. That being said, we consider the retarded self-energy to be $\Sigma^\mathrm{R}_{l\nu,l'\nu'}(\omega)=\kappa\Sigma^{\text{FBA}}_{l\nu,l\nu}\delta_{ll'}\delta_{\nu\nu'}$, where $\kappa$ represents the impurity strength.

We proceed by calculating the spectral function $A(\omega)=(-1/\pi)\text{Im}\left\{  \int dq_{1}dq_{2}\,\Tr{\Ropt{G}}  \right\}$ to see the broadening of the dispersion. The retarded Green function is $\Ropt{G}=1/(\omega+i\eta-H-\Sigma^\mathrm{R})$.
In Fig.~\ref{fig:fig2}(d), we plot the spectral function at different frequencies with different colors. Remarkably, we find much smaller lifetime for FA dispersion than the CLL dispersion, manifesting that CLL state is topologically protected as well. %This means that the FA electrons are strongly scattered, resulting in an attenuation of current contribution.
%In Fig. \ref{fig:fig3} (b), we plot spectral functions calculated with constant and gradient chiral gauge fields, in red and blue respectively. For the case with the constant $A^{5}$, the two localized Fermi arc states are both strongly affected by the impurity (red line). By contrast, for the case with the gradient $A^{5}(z)$, the broadening of the CLL state becomes smaller.

\textit{Anomalous Current and Its Fermi Level Dependency} --- 
We now focus on the two scenarios depicted in Fig.~\ref{fig:fig1}(c) and \ref{fig:fig1}(d) (no impurity), numerically calculating the current in the $y$ direction by~\cite{haug2008quantum,stefanucci2013nonequilibrium}
\begin{align}
\langle J_{y}\rangle=\frac{1}{2\pi i}
\int_{-\infty}^{\infty} d\omega \int\frac{dq_{y}dq_{z}}{(2\pi)^{2}}\,
\Tr{J_{y}G^{<}},\label{eq:Jy}
\end{align}
where $G^{<}$ is the lesser Green function, and the current operator is obtained by taking the variation of the lattice Hamiltonian $J_{y}=-\delta H/\delta A_{y}\vert_{\mathbf{A}=0}$.
In order to apply the chemical potential/temperature gradients, we utilize the standard nonequilibrium Green function technique for calculating $G^{<}=\Ropt{G}\Sigma^{<}\Aopt{G}$. The advanced Green function $\Aopt{G}$ is Hermitian adjoint of $\Ropt{G}$. The retarded self-energy $\Ropt{\Sigma}$ and the lesser self-energy $\Sigma^{<}$ describe how the system couples to fictitious electron baths which we use to implement chemical potential/temperature gradients. We consider $\Ropt{\Sigma}=-i\Gamma/2$ and $\Sigma^{<}_{ll'}=i\Gamma f(\omega,\beta_{l},\mu_{l})\delta_{ll'}$ where the standard wide-band limit approximation and a constant tunneling strength $\Gamma$ are applied. We also require independent electron baths coupled to each site, and the electron bath follows the Fermi-Dirac distribution $f(\omega,\beta_{l},\mu_{l})$ with site-dependent chemical potential $\mu_{l}$ and inverse temperature $\beta_{l}$.

In the small $\Gamma$ limit, Eq.~\eqref{eq:Jy} becomes
\begin{align}
\langle J_{y}\rangle = \frac{1}{i\Gamma} \int_{-\pi}^{\pi}\frac{dq_{y}dq_{z}}{(2\pi)^{2}}\,  \sum_{\alpha}
\langle\alpha\vert J_{y}\vert\alpha\rangle  \langle\alpha\vert\Sigma^{<}(E_{\alpha})\vert\alpha\rangle,\label{eq:Jy_nonequil}
\end{align}
where Lehmann representation $H\vert\alpha\rangle=E_{\alpha}\vert\alpha\rangle$ is applied. In Fig.~\ref{fig:fig3}(a) we show numerical current results of the chiral anomaly-like scenario by using Eq.~\eqref{eq:Jy_nonequil}: We consider a linear gradient of the chemical potential $\mu_{l}=(\Delta\mu/\Delta x)(l-L/2+1/2)+\mu_{\text{F}}$, where $\Delta\mu/\Delta x$ denotes the slope and $\mu_{\text{F}}$ represents the Fermi level, under a constant temperature $\beta_{l}=\beta$. In Fig.~\ref{fig:fig3}(b) we provide current results of the gravitational anomaly-like scenario: We consider a constant chemical potential $\mu_{l}=\mu_{\text{F}}$ but with a linear increase of the inverse temperature $\beta_{l}=\beta_{0}+(\Delta\beta/\Delta x)l/L$, where $\beta_{0}$ is the inverse temperature on the surface and $\Delta\beta/\Delta x$ represents the slope. We find that nonzero anomalous currents appear in both nonequilibrium cases, and the two scenarios exhibit distinct characteristics (odd and even functions). In fact, both signatures have been observed in the experiment~\cite{kawaguchi2020giant}, where the measurable electron density $n$ is related to the Fermi energy $\mu_{\text{F}}$ through $\mu_{\text{F}}\propto n^{2/3}$.

\textit{Transport Formula} --- 
% We then analyze the underlying mechanism of the anomalous currents.
%Notice that the lattice Hamiltonian possesses several symmetries among which we will need the particle-hole symmetry $\gamma^{3}H(\mathbf{q})(\gamma^{3})^{\dagger}=-H(-\mathbf{q})$ later.
%, $\sigma^{23}H(\mathbf{q})(\sigma^{23})^{\dagger}=H^{*}(\mathbf{q})$ and $\sigma^{02}H(q_{1},q_{2})(\sigma^{02})^{\dagger}=-H(q_{1},-q_{2})$ later (here we define $\sigma^{\mu\nu}=i[\gamma^{\mu},\gamma^{\nu}]/2$).
%\begin{figure}
%\centering
%\includegraphics[width=0.25\textwidth]{fig4}
%\caption{Calculation for current by using Eq. (\ref{eq:Jy}), manifesting the emergence of chiral anomaly.\label{fig:fig4}}
%\end{figure}
To understand the underlying mechanism of the anomalous current, we follow the derivation of the anomaly-induced transport~\cite{landsteiner2016notes}, constructing an effective two-band model of FA and CLL states to illustrate the low energy physics near the crossing points. For $\alpha=\left\{\text{FA},\text{CLL}\right\}$, $\langle\alpha\vert J_{y}\vert\alpha\rangle=\pm v$ and $E_{\alpha}=\pm vq_{y}$ in Eq.~\eqref{eq:Jy_nonequil}. For simplicity, as schematically shown in Fig.~\ref{fig:fig2}(c), we consider a localized chemical potential $\mu(x)$ and a localized inverse temperature $\beta(x)$ such that: $\left\{\mu(x),\beta(x)\right\}=\left\{\mu_{\text{FA}},\beta_{\text{FA}}\right\}$ when $x<l^{\text{FA}}$, where $l^{\text{FA}}$ is the delocalization length of the FA state, and $\left\{\mu(x),\beta(x)\right\}=\left\{\mu_{\text{CLL}},\beta_{\text{CLL}}\right\}$ elsewhere. By doing so, the matrix element of the lesser self-energy for the FA state becomes, $\langle\text{FA}\vert\Sigma^{<}\vert\text{FA}\rangle=\sum_{ll'}^{L}\langle\text{FA}\vert l\rangle \left(\Sigma^{<}\right)_{ll'} \langle l'\vert\text{FA}\rangle=i\Gamma f(E_{\text{FA}},\beta_{\text{FA}},\mu_{\text{FA}})$, where $\langle l\vert\text{FA}\rangle$ represents the FA wavefunction at the $l$-th site.
Similarly, for the CLL state, $\langle\text{CLL}\vert\Sigma^{<}\vert\text{CLL}\rangle
= i\Gamma f(E_{\text{CLL}},\beta_{\text{CLL}},\mu_{\text{CLL}})$. As a result, Eq.~\eqref{eq:Jy_nonequil} becomes
\begin{align}
\langle J_{y}\rangle=\frac{A^{5}}{2\pi^{2}}
\left(  \mu_{\text{CLL}}\tanh{\frac{\beta_{\text{CLL}}v\Lambda}{2}}  -
\mu_{\text{FA}}\tanh{\frac{\beta_{\text{FA}}v\Lambda}{2}}  \right),
\label{eq:Jy_mbg_general}
\end{align}
plus $\mathcal{O}(\mu_{\text{CLL}}^{3},\mu_{\text{FA}}^{3})$. Note that $A^{5}=-B^{5}_{y}L$, and the chiral states are estimated to exist along the $q_{z}$ coordinate within the range $[-A^{5},A^{5}]$. Additionally, since the integrand in Eq.~\eqref{eq:Jy_nonequil} does not depend on $q_{z}$, the asymmetry between the two chiral states solely comes from the integral over $q_{y}$, as depicted in Fig.~\ref{fig:fig1}(c) and \ref{fig:fig1}(d).% Note that the second order term vanishes.\\

A finite current can be obtained when either $\mu_{\text{FA}}\neq\mu_{\text{CLL}}$ or $\beta_{\text{FA}}\neq\beta_{\text{CLL}}$ (or both). On one hand, when $\beta_{\text{FA}}=\beta_{\text{CLL}}=\beta$, Eq.~\eqref{eq:Jy_mbg_general} reduces to
\begin{align}
\langle J_{y}\rangle=\frac{A^{5}}{2\pi^{2}} \tanh{\left(\frac{\beta v\Lambda}{2}\right)} (\mu_{\text{CLL}}-\mu_{\text{FA}}),\label{eq:Jy_mu_grad}
\end{align}
which shows an even function of $\mu_{\text{F}}$ as in Fig.~\ref{fig:fig3}(a) (set $\mu_{\text{FA}}=\mu_{\text{F}}-\delta$, $\mu_{\text{CLL}}=\mu_{\text{F}}+\delta$, and see $\langle J_{y}\rangle$ is independent of $\mu_{\text{F}}$). On the other hand, when $\mu_{\text{FA}}=\mu_{\text{CLL}}=\mu_{\text{F}}$, Eq.~\eqref{eq:Jy_mbg_general} becomes
\begin{align}
\langle J_{y}\rangle=\frac{A^{5}}{2\pi^{2}}
\left(  \tanh{\frac{\beta_{\text{CLL}}v\Lambda}{2}}  -
\tanh{\frac{\beta_{\text{FA}}v\Lambda}{2}}  \right)
\mu_{\text{F}},\label{eq:Jy_beta_grad}
\end{align}
which verifies the odd-function profile in Fig.~\ref{fig:fig3}(b). Most importantly, while the UV completion $\Lambda\rightarrow\infty$ is usually taken in both high energy and condensed matter literature, leading to $\langle J_{y}\rangle\rightarrow(A^{5}/2\pi^{2})(\mu_{\text{CLL}}-\mu_{\text{FA}})$ in Eq.~\eqref{eq:Jy_mu_grad} and $\langle J_{y}\rangle\rightarrow0$ in Eq.~\eqref{eq:Jy_beta_grad}, our calculations demonstrate the importance of a careful treatment for the cutoff $\Lambda$ --- We would not get the temperature dependence in Eq.~\eqref{eq:Jy_mu_grad} and a finite current in Eq.~\eqref{eq:Jy_beta_grad} if we simply take the UV limit. Note that the UV limit of Eq.~\eqref{eq:Jy_mu_grad} is the generalized CME result involving both the bulk and boundary current contributions, and the temperature dependence as well. For completeness, we also derive an analogous expression to Eq.~\eqref{eq:Jy_mbg_general} for the energy current, as details provided in \cite{SM}. There we also find two contributions to the energy current --- One from the surface FA state and the other from the bulk CLL state, and a careful treatment of the UV cutoff also reveals additional components.

Finally we investigate the temperature dependence of the lattice calculation by using Eq.~\eqref{eq:Jy_nonequil}, and check the consistency to Eq.~\eqref{eq:Jy_mu_grad}. We focus on $\mu_{\text{F}}=0$ and simulate different values of $\Delta\mu/\Delta x$ and $\beta$. Following Fig.~\ref{fig:fig2}(c), we assign the value of chemical potential at $0$th site to $\mu_{\text{FA}}$, and use $\mu_{\text{F}}$ as an approximate $\mu_{\text{CLL}}$, since the Gaussian-profile CLL state homogeneously delocalizes around the real space over different values of $q_{z}$. In Fig.~\ref{fig:fig3}(c), we find clearly linear dependency of $\langle J_{y}\rangle$ on $\mu_{\text{CLL}}-\mu_{\text{FA}}$ and the data points asymptotically approach a fixed line at low temperature. These findings are in line with Eq.~\eqref{eq:Jy_mu_grad}.

\textit{Mass dependency} --- In Fig.~\ref{fig:fig3}(d), we plot $\langle J_{y}\rangle$ at $\mu_{\text{F}}=0$ as a function of mass, ranging from massless to the massive topologically trivial phase (phase transition occurs around $m=-1.2$). In the topological phase, the current depends on the length of flat band region along the $q_{z}$ direction, corresponding to the integral over $q_{z}$ in Eq.~\eqref{eq:Jy_mbg_general}. The current decreases as $\abs{m}$ increases; however, it remains finite even in the trivial phase. The result suggests that the ISHE can be substantially enhanced in the massless case.
%(i) our model can be considered as a microscopic understanding of the ISHE, and (ii) 

\textit{Conclusion} --- We numerically and analytically demonstrate the existence of anomalous charge currents in the three dimensional Dirac fermionic systems when two crucial elements are applied: A background CGF, which we allow to be space dependent, and a thermodynamic gradient. We establish a connection between these elements and the necessary components in the ISHE. The charge currents as a function of the Fermi level are even and odd symmetric in the presence of the chemical potential gradient and the temperature gradient, respectively. Microscopically the finite anomalous currents are generated by the imbalance between the two chiral states --- the FA state localized at the boundary and the delocalized CLL state induced by the space dependency of the CGF. We numerically verify that the CLL state is actually more robust against impurity than the FA state. Additionally, we derive the transport formula Eq.~\eqref{eq:Jy_mbg_general} for the anomalous current, from which we argue that only when we carefully keep the UV cutoff can we explain the lattice calculations. A similar expression for the energy current is also provided.

Inspired by Ref.~\cite{PhysRevLett.126.216405}, we have also investigated a non-Hermitian anomaly-like scenario where we apply an inhomogeneous tunneling strength $\Gamma_{l}$ as a function of site. To our surprise, no anomalous current is observed, even with artificial negative lifetimes for partial sites. It is worth stressing that our findings align with recent experimental observations, indicating that our model can be potentially considered as a microscopic candidate for the ISHE.

 \section*{Acknowledgement}
The authors appreciate the fruitful discussions with Y.~Araki, Y.~Fuseya, M.~Hayashi, T.~Morimoto, K.~Nomura, A.~Ozawa, and G.~Tatara. This work is supported by JST CREST Grant No.~JPMJCR19T3, Japan.
S.O. is supported by JSPS KAKENHI Grants Nos.~JP22K13998 and JP23K25816.

\bibliography{main}

% \onecolumngrid
\clearpage

\renewcommand{\thesection}{\Alph{section}}
\renewcommand{\theequation}{\thesection\arabic{equation}}
\renewcommand{\thefigure}{\thesection\arabic{figure}}
\setcounter{section}{0}
\setcounter{equation}{0}
\setcounter{figure}{0}

% \begin{center}
% \bf\large Supplemental Material --- Inverse Spin Hall Effect in Nonequilibrium Dirac Systems Induced by Anomalous Flow Imbalance
% \end{center}

% \begin{center}
% Hung-Hsuan Teh\\
% \textit{Department of Chemistry, University of Pennsylvania, Philadelphia, Pennsylvania 19104, USA}
% \end{center}
%
%
%
%\section{Supplementary Material here}
%\begin{align}
%w_{p}=\prod_{j\in\mathrm{boundary}(p)}s_{j}.\label{eq:kitaev1_appendix}
%\end{align}
%\begin{figure}
%\centering
%\includegraphics[width=0.4\textwidth]{fig}
%\caption{Figure taken from Ref. ~\citenum{KITAEV20062}.}
%\end{figure}

%\begin{acknowledgments}
%\end{acknowledgments}

\appendix
\section{Chiral Landau Level State}
In this section we provide details of deriving the chiral Landau level state, namely Eq.~\eqref{eq:cll_state} in the main body of the text. When $m=0$, Eq.~\eqref{eq:bloch_H_dirac_A5} in the main body of the text becomes block diagonal $H=[H_{\text{L}}\,0;\,0\,H_{\text{R}}]$ where $\text{L}$/$\text{R}$ label left/right Weyl fermions, and $H_{\text{L},\text{R}}=\mp v\sum_{j}q_{j}\sigma^{j}+A^{5}_{z}(x)\sigma^{z}$. We first quantize the mechanical momenta in $H_{\text{L}}$ by
\begin{align*}
\pi_{x}&=-vq_{x}=\sqrt{B^{5}v/2}(a+a^{\dagger})\\
\pi_{z}&=-vq_{z}+A^{5}(1-x/\xi)=i\sqrt{B^{5}v/2}(a-a^{\dagger}),    
\end{align*}
where the ladder operators satisfy $[a,a^{\dagger}]=1$. The left Weyl Hamiltonian then becomes
\begin{align*}
H_{\text{L}}=\pi_{x}\sigma^{x}-vq_{y}\sigma^{y}+\pi_{z}\sigma^{z}.
\end{align*}
The eigenvalue problem can be solved, in a slightly easier way, by first rewriting $H_{\text{L}}$ in the eigen basis of $\sigma^{y}$: $H_{\text{L}}'=U_{y}^{\dagger}H_{\text{L}}U_{y}$ where $U_{y}=[i\, i;\, -1\, 1]/\sqrt{2}$.
%$H_{\text{L}}=[i\sqrt{B^{5}v/2}(a-a^{\dagger})\quad \sqrt{B^{5}v/2}(a+a^{\dagger})+ivq_{y};\quad \sqrt{B^{5}v/2}(a+a^{\dagger})-ivq_{y}\quad -i\sqrt{B^{5}v/2}(a-a^{\dagger})]$ where $B^{5}=A^{5}/\xi$.
%$H_{\text{L}}=v[-q_{z}\quad i\sqrt{2B^{5}}a_{\text{L}}^{\dagger};\quad -i\sqrt{2B^{5}}a_{\text{L}}\quad q_{z}]$.
%\begin{align}\label{eq:H_left_right_cll}
%H_{\text{L}}=v
%\begin{bmatrix}
%-q_{z} & i\sqrt{2B^{5}}a_{\text{L}}^{\dagger}\\
%-i\sqrt{2B^{5}}a_{\text{L}} & q_{z}
%\end{bmatrix}.
%\end{align}
$H_{\text{L}}'$ has eigenstates
\begin{align*}
\vert\psi_{n}'\rangle\sim
\begin{bmatrix}
    &u_{n}\vert n\rangle\\
    &v_{n}\vert n-1\rangle
\end{bmatrix}
\end{align*}
with bulk spectra $E_{n}=\pm\sqrt{2B^{5}vn+v^{2}q_{y}^{2}}$ for $n\geq1$ (See Fig.~\ref{fig:fig1}(b) black lines in the main body of the text). Here $u_{n},v_{n}\in\mathbb{C}$. The harmonic mode $\vert n\rangle$ satisfies $a\vert n\rangle=\sqrt{n}\vert n-1\rangle$ and $a^{\dagger}\vert n\rangle=\sqrt{n+1}\vert n+1\rangle$.

In addition, $H_{\text{L}}'$ has a $0$th chiral Landau level (CLL) eigenstate
\begin{align*}
\vert\psi_{0}'\rangle\sim
\begin{bmatrix}
    u_{0}\vert0\rangle\\
    0
\end{bmatrix}    
\end{align*}
with eigenenergy $E_{0}=-vq_{y}$ (See Fig.~\ref{fig:fig1}(b) blue line in the main body of the text). Note that the dispersion along the $q_{z}$ direction remains flat bands for the (chiral) LL states.\\
The CLL wave function in real space can be obtained by solving
\begin{align*}
&\langle x\vert a\vert\psi_{0}'\rangle\sim
\begin{bmatrix}
    \langle x\vert a u_{0}\vert0\rangle\\
    0
\end{bmatrix}
=0,
\end{align*}
meaning that
\begin{align*}
\langle x\vert a\vert0\rangle=&\sqrt{\frac{1}{2B^{5}v}}\langle x\vert\left(\pi_{x}-i\pi_{z}\right)\vert0\rangle\\
=&\sqrt{\frac{1}{2B^{5}v}}
\left[  iv\partial_{x}-iA^{5} \left(1-\frac{x}{\xi}\right) +ivq_{z}  \right]\phi_{0}^{\text{L}}(x)\\
=&0,
\end{align*}
where $\phi_{0}^{\text{L}}(x)=\langle x\vert0\rangle$. The solution is a Gaussian distribution
\begin{align*}
\phi_{0}^{\text{L}}(x)\sim  e^{-A^{5}x^{2}/2v\xi+(-q_{z}+A^{5}/v)x}.
\end{align*}
Through the similar procedures, we obtain the same bulk and chiral spectra for the right Weyl Hamiltonian $H_{\text{R}}$, though the corresponding Gaussian distribution has a slightly different center:
\begin{align*}
\phi_{0}^{\text{R}}(x)\sim
e^{-A^{5}x^{2}/2v\xi + (q_{z}+A^{5}/v)x}.
\end{align*}
As a result, we can combine the solutions for $H_{\text{L}/\text{R}}$ to obtain the $0$th CLL state of the total Hamiltonian $H$. After transforming back to the eigen basis of $\sigma^{z}$, we get Eq.~\eqref{eq:cll_state} in the main body of the text.

\section{Energy Current}
In this section we derive the energy current, which is analogous to the charge current in Eq.~\eqref{eq:Jy_mbg_general} in the main body of the text, for the effective two band model of the FA and CLL states. The derivation follows similar procedure to achieve Eq.~\eqref{eq:Jy_mbg_general} in the main text, except that the charge current operator $J_{y}$ is replaced by the energy current operator $J^{\text{E}}_{y}$,
\begin{align*}
\langle J^{\text{E}}_{y}\rangle = \frac{1}{i\Gamma} \int_{-\pi}^{\pi}\frac{dq_{y}dq_{z}}{(2\pi)^{2}}\,  \sum_{\alpha}
\langle\alpha\vert J^{\text{E}}_{y}\vert\alpha\rangle  \langle\alpha\vert\Sigma^{<}(E_{\alpha})\vert\alpha\rangle.
\end{align*}
In relativistic theory, $\langle\alpha\vert J^{\text{E}}_{y}\vert\alpha\rangle=q_{y}$ for $\alpha=\{\text{FA},\text{CLL}\}$, and therefore,
\begin{align*}
\langle J^{\text{E}}_{y}\rangle=&
\frac{A^{5}}{2\pi^{2}}\int_{-\Lambda}^{\Lambda}dq_{y}\Big\{
q_{y}f(E_{\text{FA}},\beta_{\text{FA}},\mu_{\text{FA}})\notag\\
&+q_{y}f(E_{\text{CLL}},\beta_{\text{CLL}},\mu_{\text{CLL}})
\Big\}\notag\\
=&\frac{A^{5}}{2\pi^{2}v^{2}}\int_{-v\Lambda}^{v\Lambda}dx\,
\left\{
\frac{x}{e^{\beta_{\text{FA}}(x-\mu_{\text{FA}})}+1}
-\frac{x}{e^{\beta_{\text{CLL}}(x-\mu_{\text{CLL}})}+1}
\right\}.
\end{align*}
The integral has the following closed form,
\begin{align}
\langle J^{\text{E}}_{y}\rangle  =  \frac{A^{5}}{2\pi^{2}v^{2}}
\Big[\mathcal{I}\left(\beta_{\text{FA}},\mu_{\text{FA}}\right)
-\mathcal{I}\left(\beta_{\text{CLL}},\mu_{\text{CLL}}\right)\Big],\label{eq:JEy_mgb_general}
\end{align}
where
\begin{align}
\mathcal{I}(\beta,\mu)\equiv&
-\frac{v\Lambda}{\beta}  \ln{\left[\left(1+e^{-\beta v\Lambda}\right) \left(1+e^{\beta v\Lambda}\right)\right]}\notag\\
&+\frac{1}{\beta^{2}}\left[
\text{Li}_{2}\left(-e^{-\beta v\Lambda}\right)  -\text{Li}_{2}\left(-e^{\beta v\Lambda}\right)
\right]\notag\\
&+\left[
-\frac{\beta v\Lambda e^{\beta v\Lambda}}{\left(e^{\beta v\Lambda}+1\right)^{2}}
-\frac{1}{e^{\beta v\Lambda}+1}
+\frac{1}{2}
\right]\mu^{2}
+\mathcal{O}(\mu^{3}).\label{eq:JEy_I}
\end{align}
Here $\text{Li}_{2}$ denotes the dilogarithm function. Equations (\ref{eq:JEy_mgb_general}) and (\ref{eq:JEy_I}) provide the general energy current expressions, which involve not only $\beta$ and $\mu$ but also the UV cutoff $\Lambda$. In the UV limit where $\Lambda\rightarrow\infty$, we recover the chiral magnetic effect in the energy current \cite{Loganayagam:2012pz,landsteiner2016notes}:
\begin{align*}
\langle J^{\text{E}}_{y}\rangle=
\frac{A^{5}}{12v^{2}}\left(  \frac{1}{\beta_{\text{FA}}^{2}}  -\frac{1}{\beta_{\text{CLL}}^{2}}  \right)
+\frac{A^{5}}{4v^{2}\pi^{2}}\left(  \mu_{\text{FA}}^{2}  -\mu_{\text{CLL}}^{2}  \right).
\end{align*}
However, as demonstrated for the charge current in the main body of the text, the cutoff dependency is crucial for explaining the lattice calculations, so that the extra contribution must be carefully included.

\end{document}